\documentclass[prl,aps,twocolumn,showpacs,preprintnumbers,%
amsmath,amssymb]{revtex4}

\usepackage{graphicx}	

\begin{document}

\title{Flavor oscillations in the supernova hot bubble region:\\
Nonlinear effects of neutrino background}
\author{Sergio Pastor and Georg Raffelt}
\affiliation{Max-Planck-Institut f\"ur Physik
(Werner-Heisenberg-Institut), F\"ohringer Ring 6, 80805 Munich,
Germany}

\date{October 11, 2002}

\begin{abstract}
The neutrino flux close to a supernova core contributes substantially
to neutrino refraction so that flavor oscillations become a nonlinear
phenomenon. One unexpected consequence is efficient flavor
transformation for anti-neutrinos in a region where only neutrinos
encounter an MSW resonance or vice versa.  Contrary to previous
studies we find that in the neutrino-driven wind the electron fraction
$Y_e$ always stays below~0.5, corresponding to a neutron-rich
environment as required by r-process nucleosynthesis. The relevant
range of masses and mixing angles includes the region indicated by
LSND, but not the atmospheric or solar oscillation parameters.
\end{abstract}

\pacs{14.60.Pq, 97.10.Cv, 97.60.Bw}

\maketitle


{\it Introduction.}--- The evidence for flavor oscillations of solar
and atmospheric neutrinos and in the LSND experiment implies mass
differences so small that refractive effects influence or even
dominate neutrino oscillations in many situations of practical
interest. However, there are only two examples where neutrinos
themselves as a medium modify the oscillations. One is the early
universe~\cite{Notzold:1987ik}, the other core-collapse
supernovae (SNe)~\cite{Mayle:1986ic}.

In a seminal paper Pantaleone~\cite{Pantaleone:eq} showed that
neutrinos as a background medium differ markedly from other
fermions. A given background neutrino may be a coherent superposition
of flavor states, implying an ``off-diagonal refractive index'' in
flavor space. The oscillations of the entire ensemble thus become a
nonlinear phenomenon with unexpected consequences. When the neutrinos
themselves dominate as a background medium, the oscillations become
``synchronized,'' i.e.\ all modes oscillate collectively with the same
frequency, a behavior first discovered by
Samuel~\cite{Samuel:1993uw}. With our collaborators we recently found
a simple physical interpretation of this perplexing phenomenon in
terms of the dipole-dipole coupling of a collection of magnetic
dipoles which spin-precess in an external magnetic
field~\cite{Pastor:2001iu}.

The first environment where $\nu$-$\nu$ refraction plays a crucial
role is the epoch of the early universe that precedes big-bang
nucleosynthesis.  If initially large flavor-dependent
$\nu$-$\bar\nu$-asymmetries exist, they may be equilibrated by
oscillations and collisions before weak-interaction freeze-out so that
the primordial helium abundance implies stringent limits on the
overall cosmic neutrino density
\cite{Lunardini:2000fy,Dolgov:2002ab,Wong:2002fa,Abazajian:2002qx}.
Depending on initial conditions the modification of the flavor
relaxation process caused by the synchronization effect is only mild,
or it may even prevent equilibrium entirely because the synchronized
oscillation frequency can become arbitrarily small.  The interplay of
simultaneous $\nu$ and $\bar\nu$ oscillations is a crucial and
non-trivial ingredient in the evolution of this system.

The second system where background neutrinos may be important is the
rarefied region just outside the nascent neutron star a few seconds
after SN core bounce.  In the innermost regions of the SN, the
neutrino flux is so large that the weak-interaction potential created
by the neutrinos is comparable to that of the ordinary medium.  The
neutrino spectra and fluxes differ between the flavors and between
neutrinos and anti-neutrinos of a given flavor.  Swapping the fluxes
of different flavors would crucially modify the production of heavy
elements via r-process nucleosynthesis if this phenomenon takes place
in the SN hot bubble region~\cite{Qian:dg,Qian:2000nf}.  The possible
importance of $\nu$-$\nu$-refraction in this context was quickly
recognized~\cite{Pantaleone:1994ns,Qian:wh,Sigl:1994hc}.  The main
consequence implied by these approximate treatments was a small shift
of the oscillation parameters where a significant spectral swapping by
resonant oscillations takes place.

Alerted by the subtleties encountered in our study of early-universe
oscillations \cite{Dolgov:2002ab} we revisit the $\nu$-$\nu$-effect in
the SN hot bubble region. We find that previous authors indeed
underestimated the complications that arise, in particular, when
neutrinos and anti-neutrinos oscillate simultaneously and cause
refractive effects for each other. We find, for example, that in a
region of parameters where neutrinos encounter an MSW resonance, the
anti-neutrinos are ``dragged along'' and also show large flavor
transformations. The final picture of the interplay between neutrino
oscillations and r-process nucleosynthesis is very different than
previously imagined.

{\it Two-Flavor System.}---To be specific we study the
\hbox{$\nu_e$-$\nu_\mu$} system with oscillation parameters $\tan^2
\theta$ and $\Delta m^2 = m^2_2-m^2_1> 0$.  In the absence of neutrino
background effects, neutrinos (anti-neutrinos) encounter an MSW
resonance for $\tan^2 \theta < 1$ ($\tan^2 \theta > 1$). The evolution
of the neutrino system is described by the $2 {\times} 2$ density
matrices
\begin{eqnarray}
\rho_{\bf p}(t) = \left (\begin{array}{cc}
\rho_{ee}& \rho_{e\mu}\\
\rho_{\mu e}& \rho_{\mu \mu} 
\end{array}
\right) =
\frac{1}{2} \left[P_0({\bf p},t) +\hbox{\boldmath$\sigma$}
\cdot{\bf P}_{\bf p}(t)\right]~,
\label{2by2mutau} 
\end{eqnarray}
and analogously $\bar{\rho}_{\bf p}$ for anti-neutrinos. Here,
${\sigma}_i$ are the Pauli matrices while ${\bf P}_{\bf p}(t)$ and
$\overline{\bf P}_{\bf p}(t)$ are the usual polarization vectors for
$\nu$ and $\bar\nu$ modes with momentum ${\bf p}$, respectively. The
diagonal elements $\rho_{\alpha\alpha}({\bf p},t)$ are the occupation
numbers of flavor $\alpha$ with momentum ${\bf p}$.

In the region of interest neutrinos stream freely so that we may
ignore collisions.  Therefore, the radial evolution equation is the
usual precession formula, augmented by the $\nu$-$\nu$ refractive term
\cite{Sigl:1993fn}
\begin{eqnarray}
\partial_r 
\begin{pmatrix}
{\bf P_p}\\
\noalign{\medskip}
{\bf \overline{P}_p}
\end{pmatrix}
&=&\Biggl\{\sqrt2\,G_{\rm F}\Biggl[N_{e}\,\hat{\bf z}+
\int d{\bf q}~C_{\bf pq}
\Bigl({\bf P_q} - {\bf \overline{P}_q}\Bigr)
\Biggr]\nonumber\\
&&\kern1em{}\pm\frac{\Delta m^2}{2p}\,{\bf B}\Biggr\}
\times 
\begin{pmatrix}
{\bf P_p}\\
\noalign{\medskip}
{\bf \overline{P}_p}
\end{pmatrix}~.
\label{polvec}
\end{eqnarray}
Here ${\bf B}=(\sin2\theta,0,-\cos2\theta)$ is a ``magnetic field,''
$\theta$ the vacuum mixing angle, and $\hat {\bf z}$ a unit vector in
the $z$-direc\-tion in flavor space.  Further, $N_{e}=Y_e N_B$ is the
electron density with $Y_e$ the electron fraction and $N_B$ the baryon
density.  Finally, $C_{\bf pq}\equiv1-{\bf \hat p}\cdot{\bf \hat q}$,
implying that collinear neutrinos do not cause a mutual refraction
effect.

As a matter density profile for the hot bubble region we use the one
shown in Ref.~\cite{Qian:dg} which roughly falls off as $r^{-3}$.  As
a boundary condition we assume equal luminosities $L_{\nu}$ for all
flavors of order $L_0\equiv 10^{51}~\rm erg~s^{-1}$.  The spectra are
taken to be Fermi-Dirac distributions with mean energies $\langle
E_{\nu_e}\rangle = 11~{\rm MeV}$, $\langle E_{\bar{\nu}_e}\rangle =
16~{\rm MeV}$ and $\langle E_{\nu_\mu,\bar{\nu}_\mu}\rangle = 25~{\rm
MeV}$, respectively.  These choices may not be entirely
realistic~\cite{Keil}, but for consistency with previous work we stick
with these traditional assumptions.

In the absence of oscillations and for radially moving neutrinos the
diagonal elements of the density matrix at radius $r$ and for neutrino
momentum $p$ are
\begin{equation}
\rho_{\alpha\alpha}(p,r)=\frac{L_{\nu}}{4\pi r^2}\,
\frac{120}{7\pi^4~T_{\nu_\alpha}^4}\,
\frac{p^2}{\exp(p/T_{\nu_\alpha})+1}
\label{nuspectra}
\end{equation}
where $T_{\nu_\alpha}=\langle E_{\nu_\alpha}\rangle/3.151$. For the
$\nu$-$\nu$ refractive effect the angular divergence of the neutrinos
is crucial. As in previous works~\cite{Qian:wh,Sigl:1994hc} we use a
flux-averaged value, i.e.\ in Eq.~(\ref{polvec}) we substitute
\begin{equation}
\int d{\bf q}\, C_{\bf pq}
({\bf P}_{\bf q}-{\bf\overline P}_{\bf q})
\times {\bf P}_{\bf p}
\to F(r)\,
({\bf P}-{\bf\overline P})\times{\bf P}_{\bf p}\,.
\end{equation}
Here, ${\bf P}$ and ${\bf\overline P}$ are the total polarization
vectors and $F(r)=\frac{1}{2}[1-(1-R_\nu^2/r^2)^{1/2}]$ is a
geometrical factor with $R_\nu$ the neutrino-sphere radius (see
\cite{Qian:wh} for a more detailed discussion of the geometrical
dependence).  Both $F(r)$ and the luminosity fall off as $r^{-2}$ so
that the $\nu$-$\nu$ refractive term scales as $r^{-4}$ at large
$r$. In the neutrino-driven wind phase the medium density typically
falls off as $r^{-3}$ so that at large distances the ordinary medium
dominates.  However, at distances of 15--30~km the neutrinos may
dominate.

{\it R-Process Nucleosynthesis}.---A key necessary condition for this
process to occur in the SN hot bubble region is that the environment
must be neutron-rich.  The neutron-to-proton ratio is fixed by the
$\beta$ processes $\nu_e + n \leftrightarrow p+e^-$ and $\bar{\nu}_e +
p \leftrightarrow n+e^+$ while charge neutrality requires
$n/p=1/Y_e-1$ \cite{Qian:dg}.  Therefore, a minimal requirement is
$Y_e<0.5$, but a successful r-process may require $Y_e\alt 0.45$.
Near weak-interaction freeze-out (WFO), at a radius 30--35~km, only
the direct $\beta$ processes are important and the electron fraction
is
\begin{equation} 
Y_e \approx \left( 1+\frac{L({\bar{\nu}_e})\bar{\epsilon}}
{L({\nu_e})\epsilon}\right )^{-1}
\label{ye2nd}
\end{equation}
where $\epsilon \equiv \langle E_{\nu_e}^2\rangle/ \langle
E_{\nu_e}\rangle$, and $\bar{\epsilon}$ the analogue for
$\bar{\nu}_e$. We take the $\nu_e$ and $\bar{\nu}_e$ cross sections on
nucleons to be equal, see however
\cite{Horowitz:1999fe,Horowitz:2001yv}.  In the absence of neutrino
oscillations and with our choice of neutrino flux parameters one finds
$Y_e \simeq 0.41$, allowing for a successful r-process.

\begin{figure}[b!]
\begin{center}
\includegraphics[width=0.75\columnwidth]{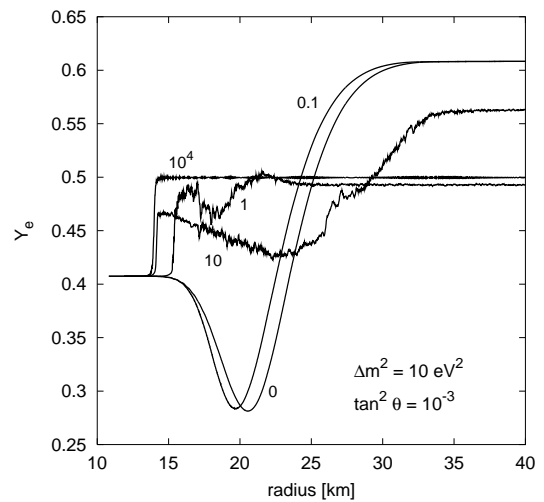}
\end{center}
\caption{$Y_e$ as a function of radius for the indicated choice of
oscillation parameters. The labels indicate $L_\nu$ in units of
$L_0=10^{51}~\rm erg~s^{-1}$; $L_\nu=0$ implies the absence of
neutrino background effects.
\label{yeevol}}
\end{figure}

{\it Spectral Swapping by Oscillations.}---If neutrino oscillations
occur within the WFO radius, the effective $\nu_e$ and $\bar\nu_e$
flux spectra change and modify $Y_e$. As a first example we use
$\Delta m^2=10~\rm eV^2$ and $\tan^2\theta=10^{-3}$ which yield the
$Y_e$ profile shown in Fig.~\ref{yeevol}.  For the curve marked $0$,
neutrino background effects were ignored, the other curves are for the
indicated values of $L_\nu$.

\begin{figure*}[htb!]
\begin{center}
\includegraphics[width=.328\textwidth]{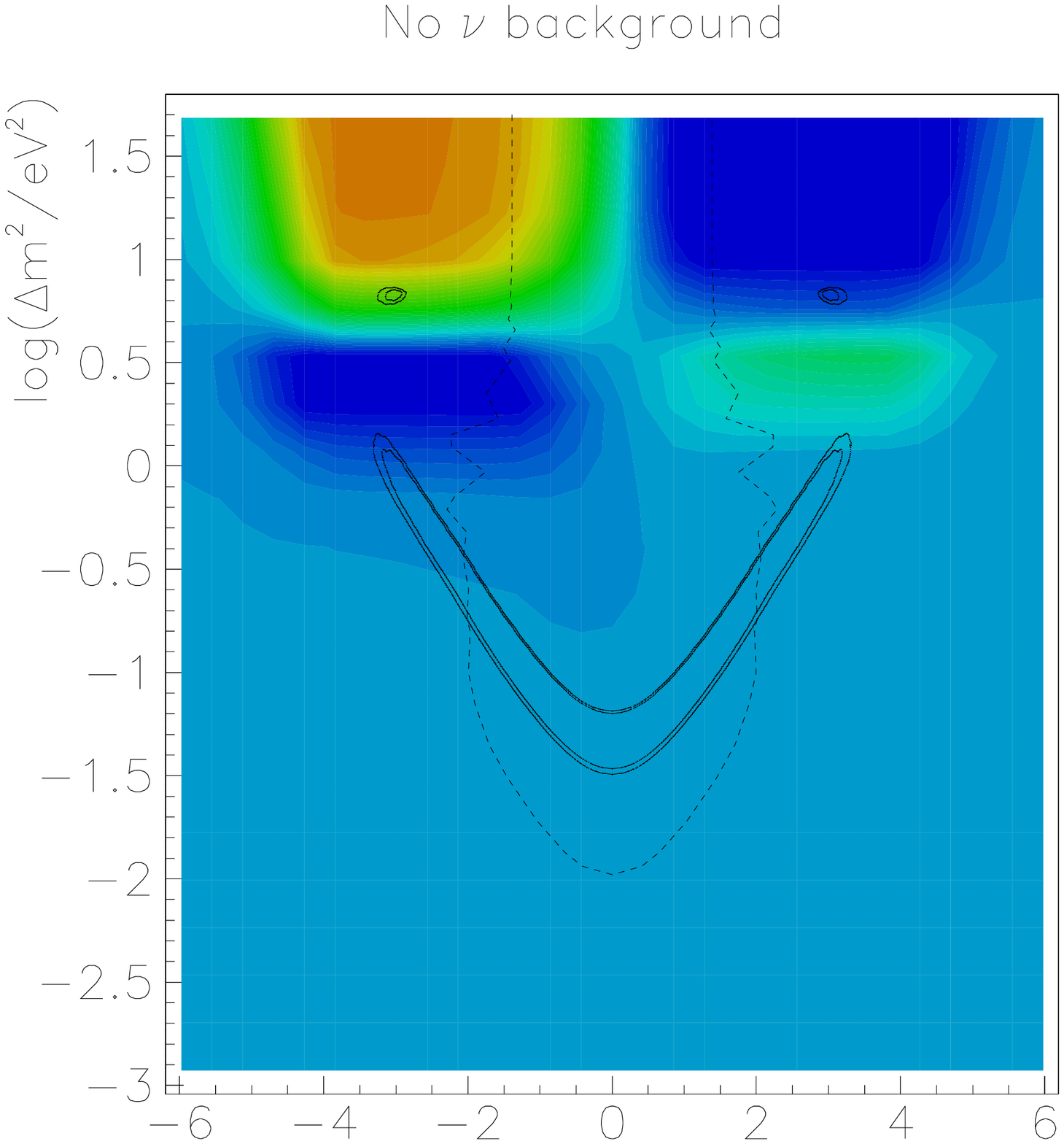}
\includegraphics[width=.328\textwidth]{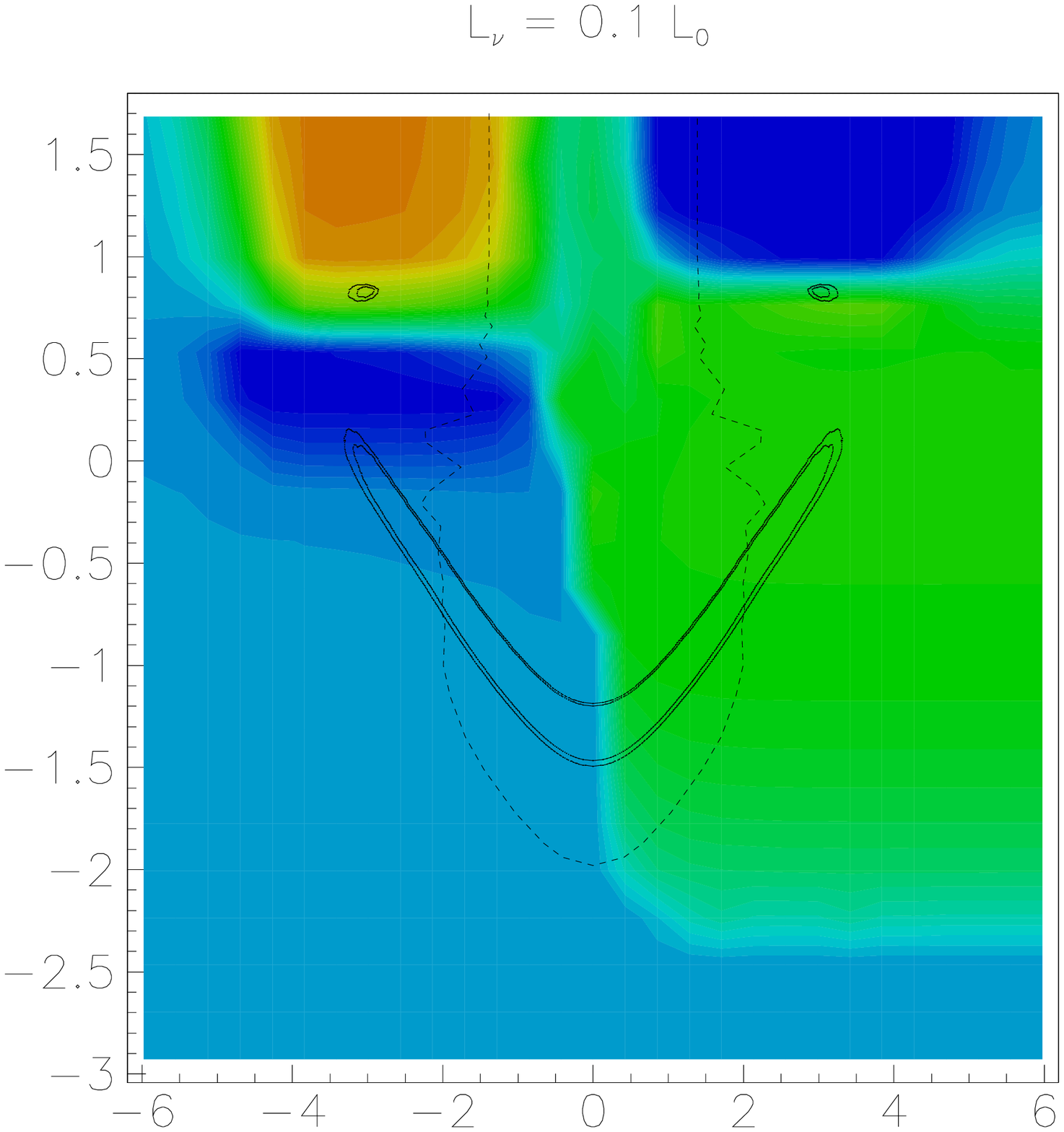}
\includegraphics[width=.328\textwidth]{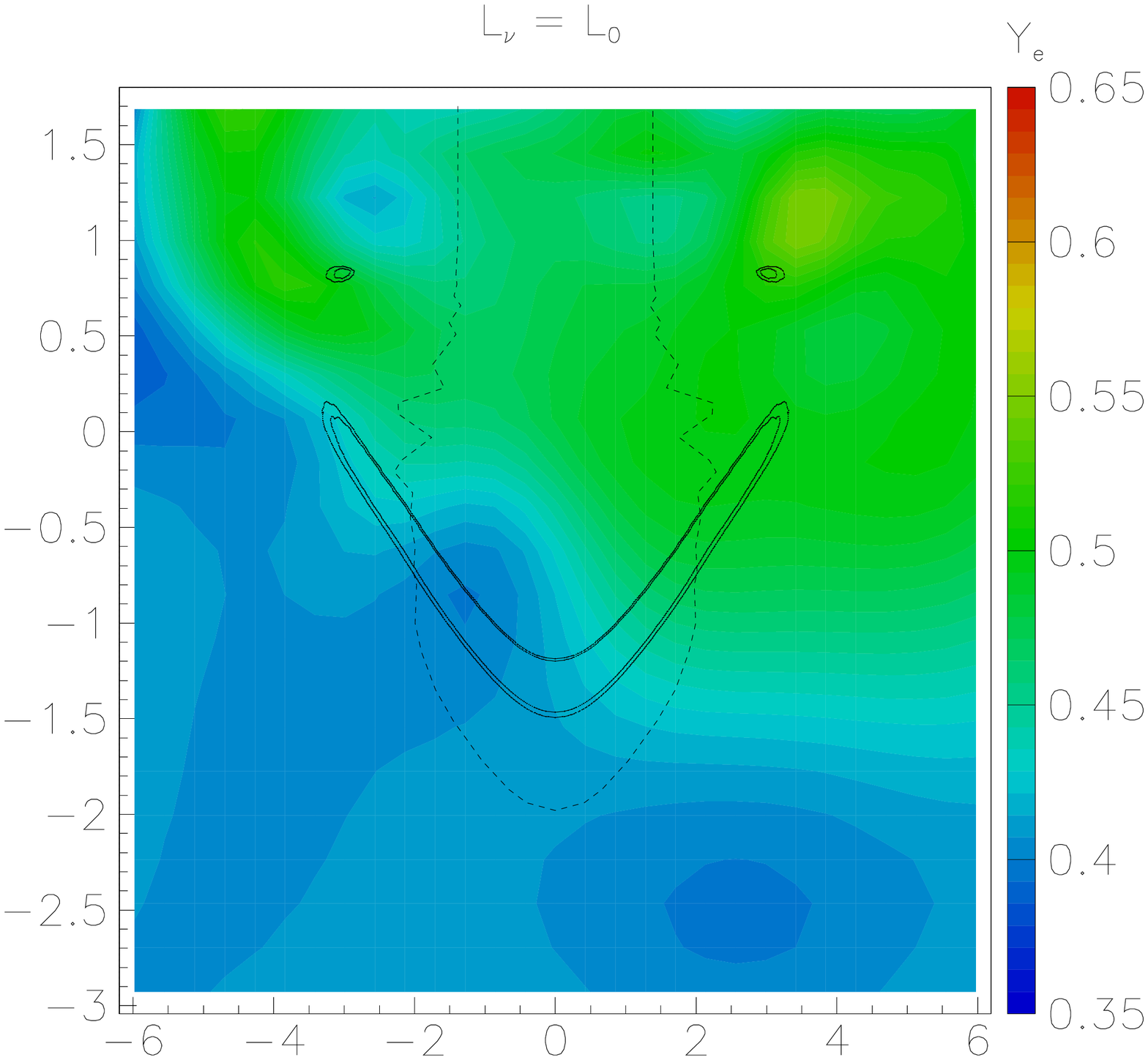}
\includegraphics[width=.328\textwidth]{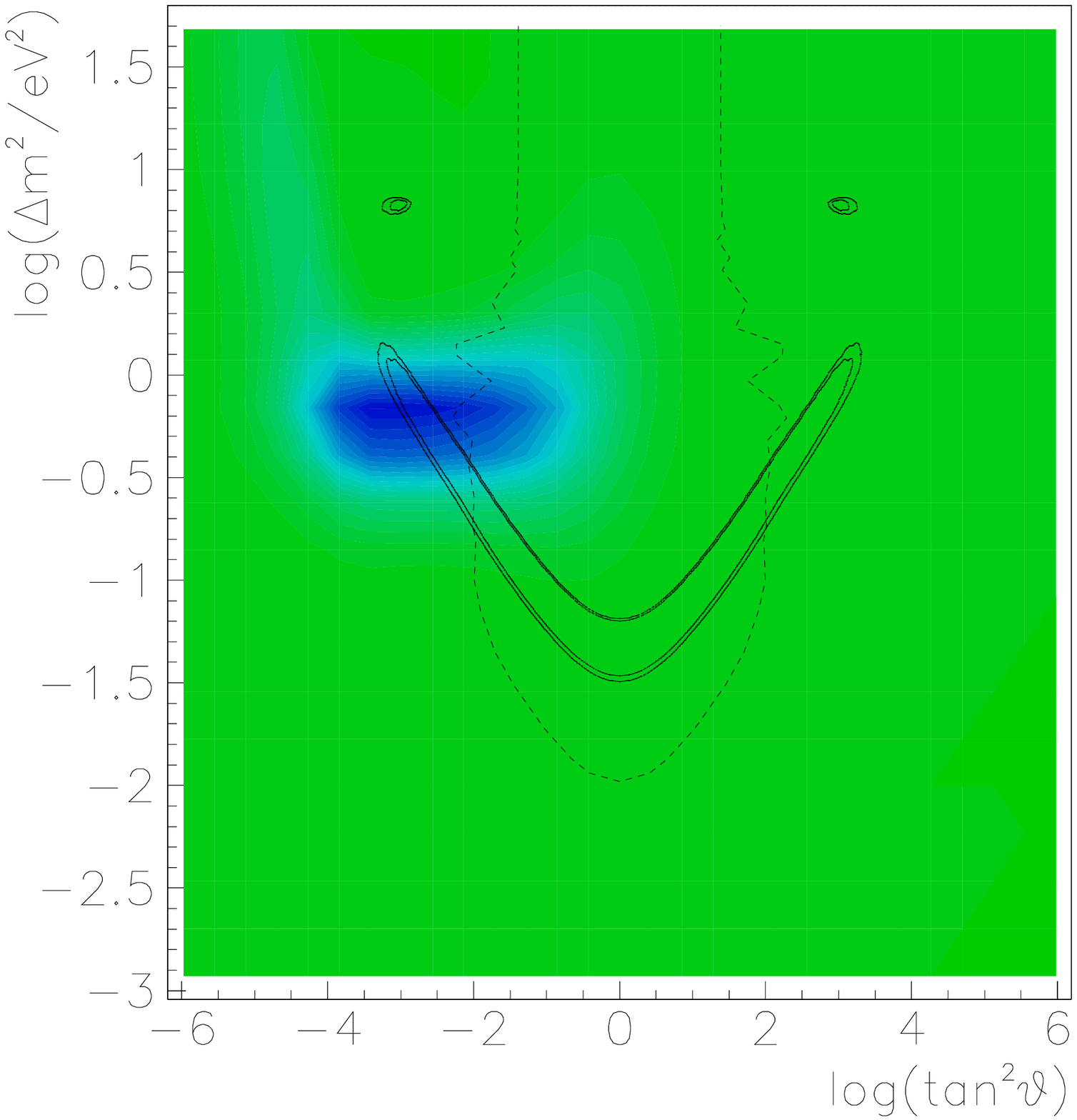}
\includegraphics[width=.328\textwidth]{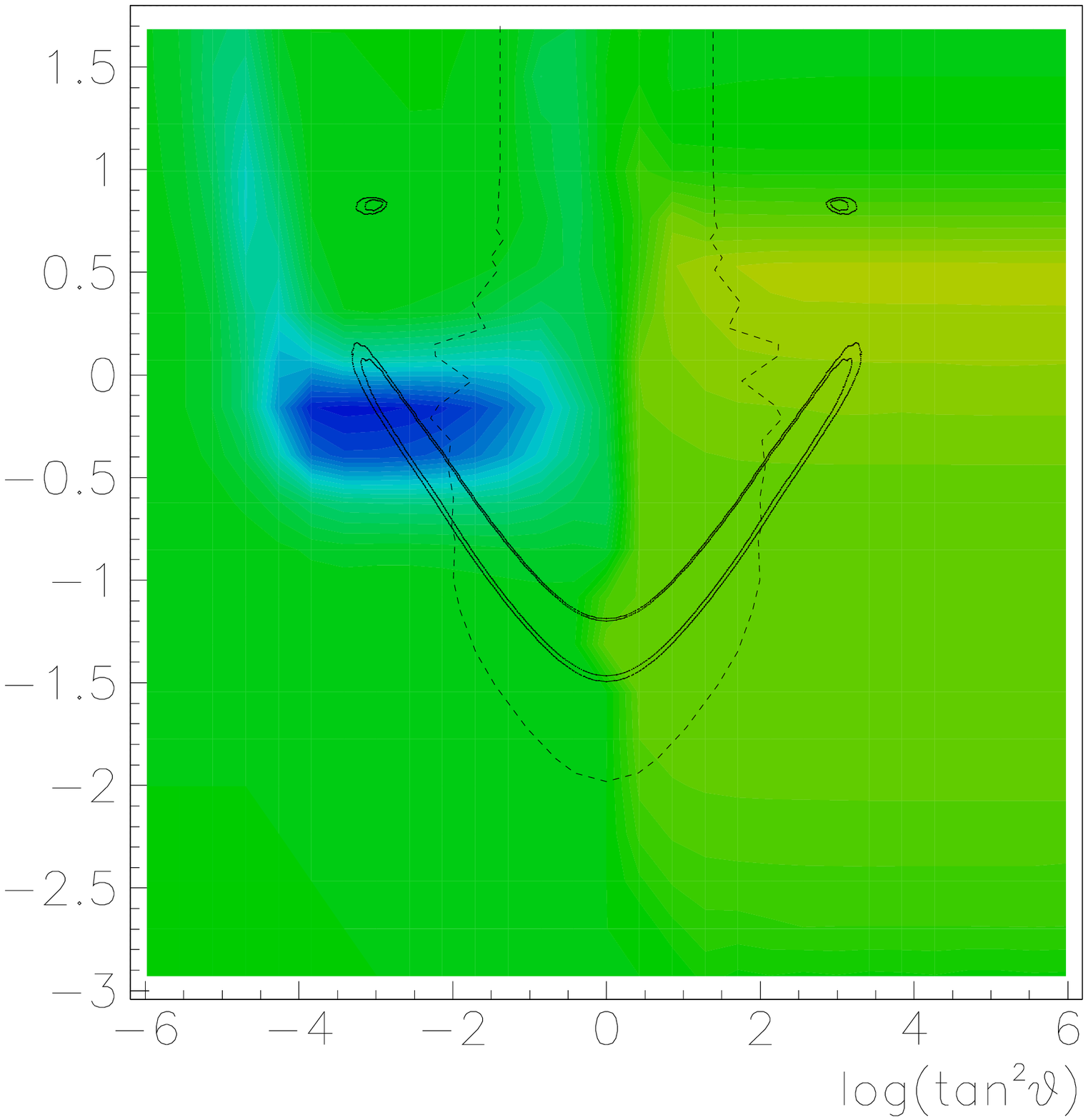}
\includegraphics[width=.328\textwidth]{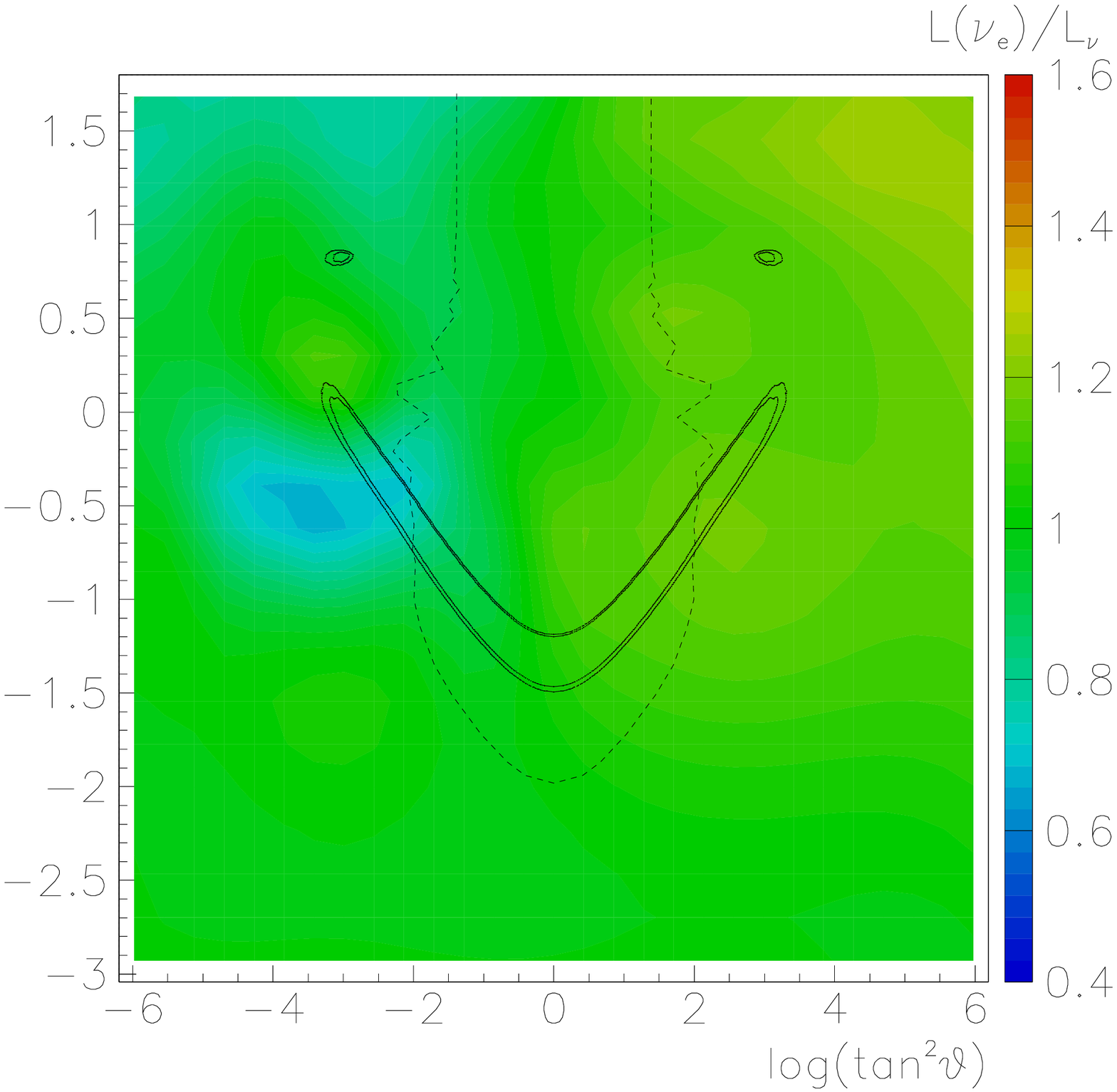}
\end{center}
\caption{Spectral swapping as a function of $\Delta m^2$ and
$\tan^2\theta$ for $L_\nu=0$ (no neutrino background effects),
$0.1$, and $1\,L_0$.  The solid contours
indicate the LSND+KARMEN2 allowed region \cite{Church:2002tc}, while
the region inside the dashed contour is excluded by the Bugey
experiment \cite{Bugey95}.  {\it Upper Panels:\/} $Y_e$~at the WFO
radius of $\simeq 30$~km.  {\it Lower Panels:\/} $L(\nu_e)/L_\nu$ at
$r=50$~km.
\label{contye}}
\end{figure*}

The oscillations can be calculated analytically in the limit $L_\nu\gg
L_0$ where the neutrino background strongly dominates.  We define
${\bf I} \equiv {\bf P} -\overline{\bf P}$, integrate
Eq.~(\ref{polvec}) over the neutrino spectra to get the evolution
equations for ${\bf P}$ and $\overline{\bf P}$, and subtract them to
obtain
\begin{equation} 
\partial_r {\bf I}= 
\int d{\bf p}\,\frac{\Delta m^2}{2p}\,{\bf B} 
\times\left[ {\bf P}_{\bf p}+\overline{\bf P}_{\bf p}\right]
+\sqrt2 \,G_{\rm F} N_e\,\hat{\bf z}
\times  {\bf I}\,.
\label{evolI}
\end{equation}
The neutrino background term is proportional to ${\bf I}\times {\bf
I}$ and thus vanishes.  However, the individual modes ${\bf P}_{\bf
p}$ and $\overline{\bf P}_{\bf p}$ precess fast around ${\bf I}$ as in
Ref.~\cite{Pastor:2001iu}.  The evolution of ${\bf I}$ is a slow
precession with a certain synchronized frequency $\omega_{\rm synch}$.
We express $\omega_{\rm synch}$ by the neutrino momentum $p_{\rm
synch}$ that would precess with $\omega_{\rm synch}$ in the absence of
a neutrino background. Note that the synchronization of both neutrino
and anti-neutrino modes occurs despite the presence of a CP asymmetric
background \cite{Wong:2002fa}.

To find $p_{\rm synch}$ we use that in the present limit all ${\bf
P}_{\bf p}$ and $\overline{\bf P}_{\bf p}$ are essentially aligned
with ${\bf I}$ so that their projections along ${\bf I}$ are
conserved.  All modes start in the $z$-direction so that altogether
${\bf P}_{\bf p}\simeq P_z({\bf p})\,{\bf\hat I}$ and ${\bf\overline
P}_{\bf p}\simeq {\overline P}_z({\bf p})\,{\bf\hat I}$.  We can then
rewrite Eq.~(\ref{evolI}) as
\begin{equation} 
\partial_r {\bf I}= 
\left[\frac{\Delta m^2}{2 p_{\rm synch}}{\bf B} 
+\sqrt2 \,G_{\rm F} N_{e}\,\hat{\bf z}\right]
\times  {\bf I}\,,
\label{evolIsync}
\end{equation}
where
\begin{equation}
p_{\rm synch} \simeq \frac{18\,\zeta_3}{\pi^2}\,
\frac
{\left[
T_{\nu_e}^{-1}-
T_{\bar{\nu}_e}^{-1}
\right]}{\left[
T_{\nu_e}^{-2}+
T_{\bar{\nu}_e}^{-2}
-2\,T_{\nu_\mu}^{-2}
\right]}\,.
\end{equation}
We have used that initially $T_{\nu_\mu}=T_{\bar{\nu}_\mu}$.  For our
assumed spectra we find $p_{\rm synch}\simeq 2.2~\rm MeV$, much
smaller than typical energies of the neutrino spectra.  As a
consequence, neutrino oscillations are effective at smaller radii and
for smaller $\Delta m^2$ than without a neutrino background, an effect
already observed in Refs.~\cite{Qian:wh,Sigl:1994hc}.

The results of Fig.~\ref{yeevol} are now easily explained in two
limiting cases.  Without neutrino background all neutrino modes
experience an independent adiabatic MSW transition starting at low
energies ($Y_e$ decreases) until the entire $\nu_e$ spectrum is
swapped with that of $\nu_\mu$, leading to the asymptotic value $Y_e
\simeq (1+\langle E^0_{\bar{\nu}_e}\rangle/\langle
E^0_{\nu_\mu}\rangle)^{-1} \simeq 0.61$.

The other limiting case ($L_\nu\gg L_0$) corresponds to a synchronized
MSW transition of the entire neutrino and anti-neutrino ensemble,
where all modes follow an adiabatic transition at the same radius
where a neutrino with momentum $p_{\rm synch}$ would do an MSW
transition in the absence of background neutrinos.  $Y_e$ takes on the
value $0.5$ because both $\nu_e$ and $\bar{\nu}_e$ are swapped with
$\nu_\mu$ and $\bar{\nu}_\mu$, respectively, and thus take on
identical spectra.

For the intermediate cases there is some degree of synchronization,
but it is gradually lost at larger radii with the dilution of the
neutrino flux. Still for the nominal neutrino luminosity $L_\nu=L_0$
the evolution is quite different from the no-background case.

\begin{figure*}[htb!]
\begin{center}
\includegraphics[width=.328\textwidth]{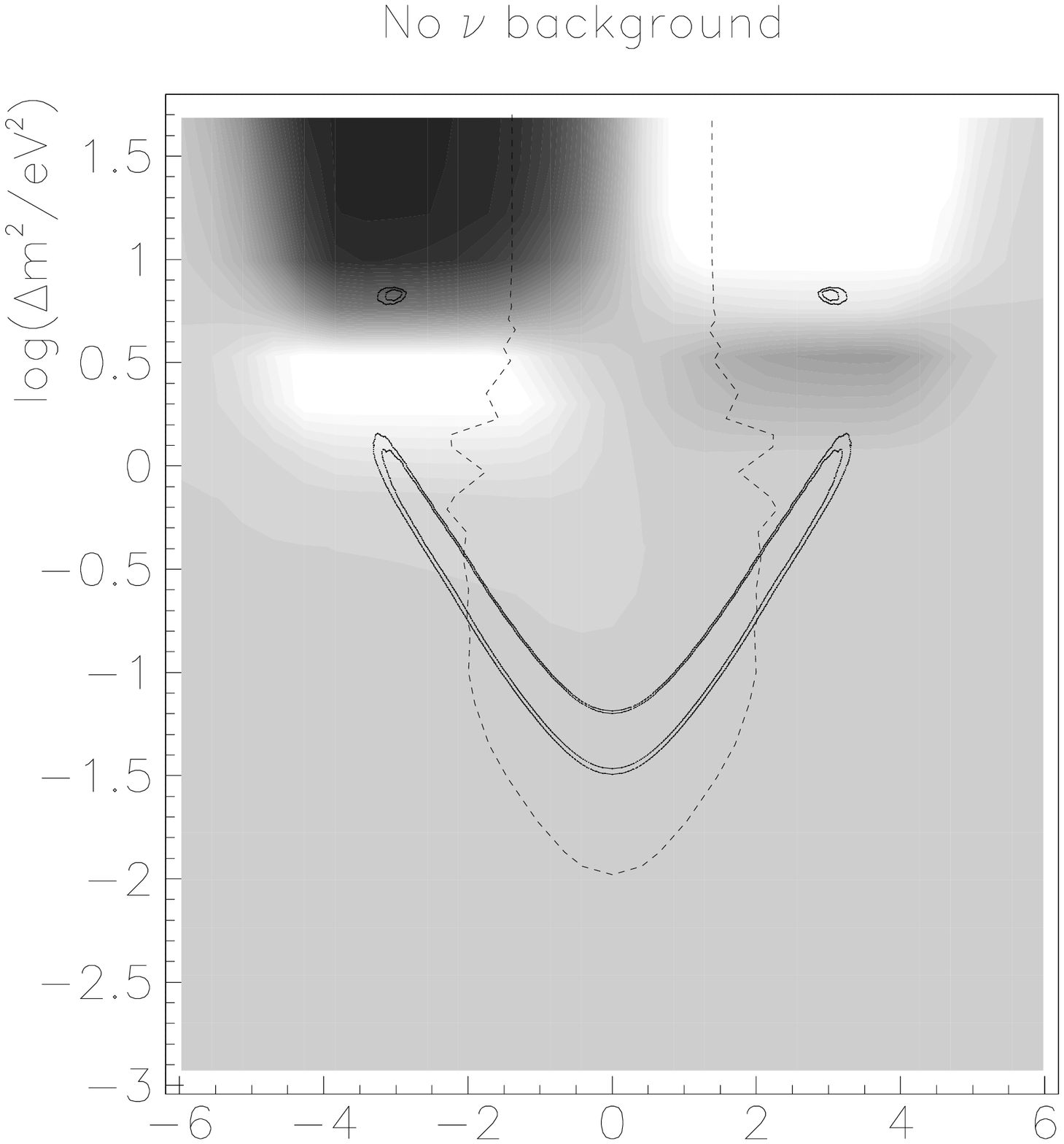}
\includegraphics[width=.328\textwidth]{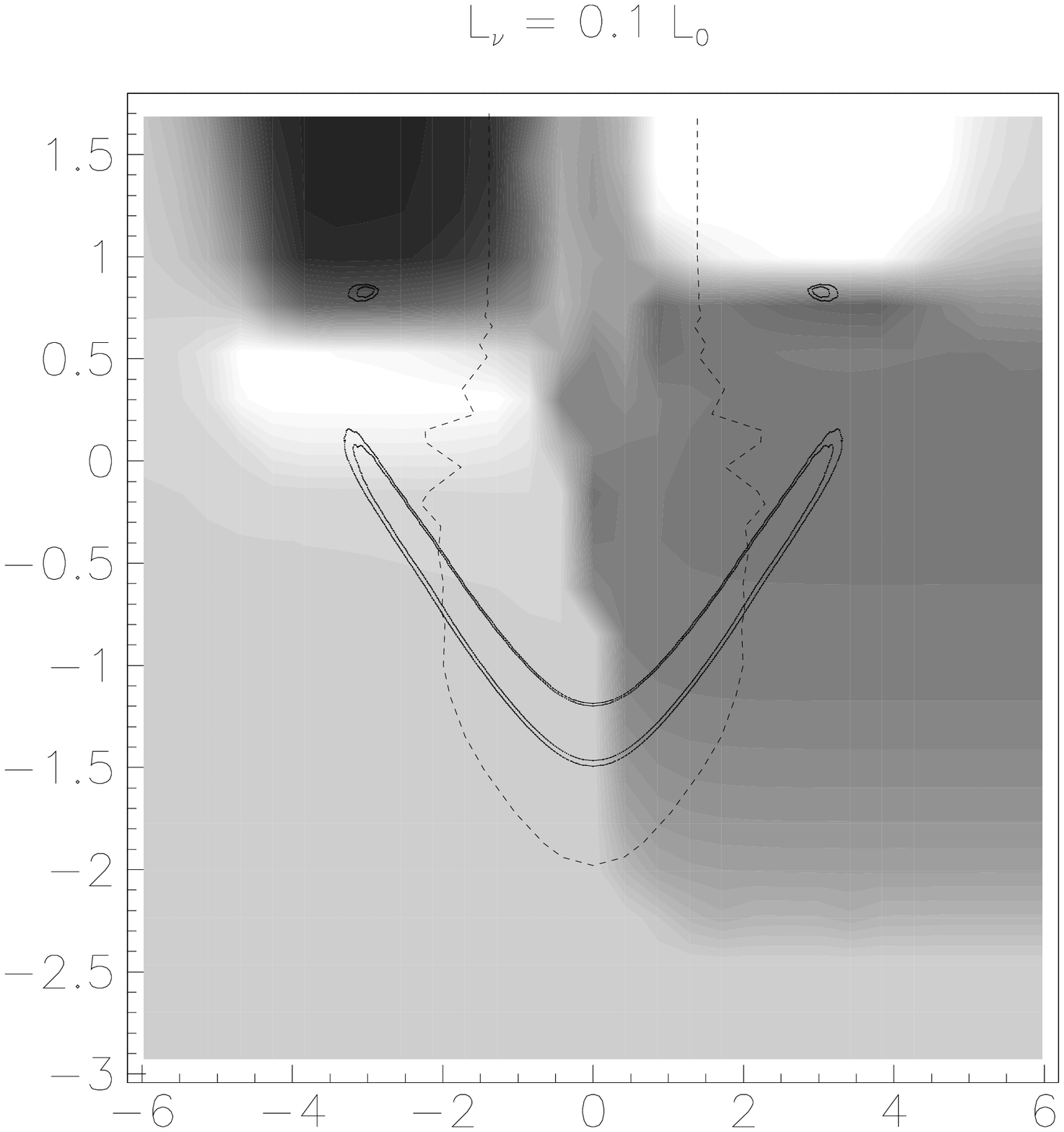}
\includegraphics[width=.328\textwidth]{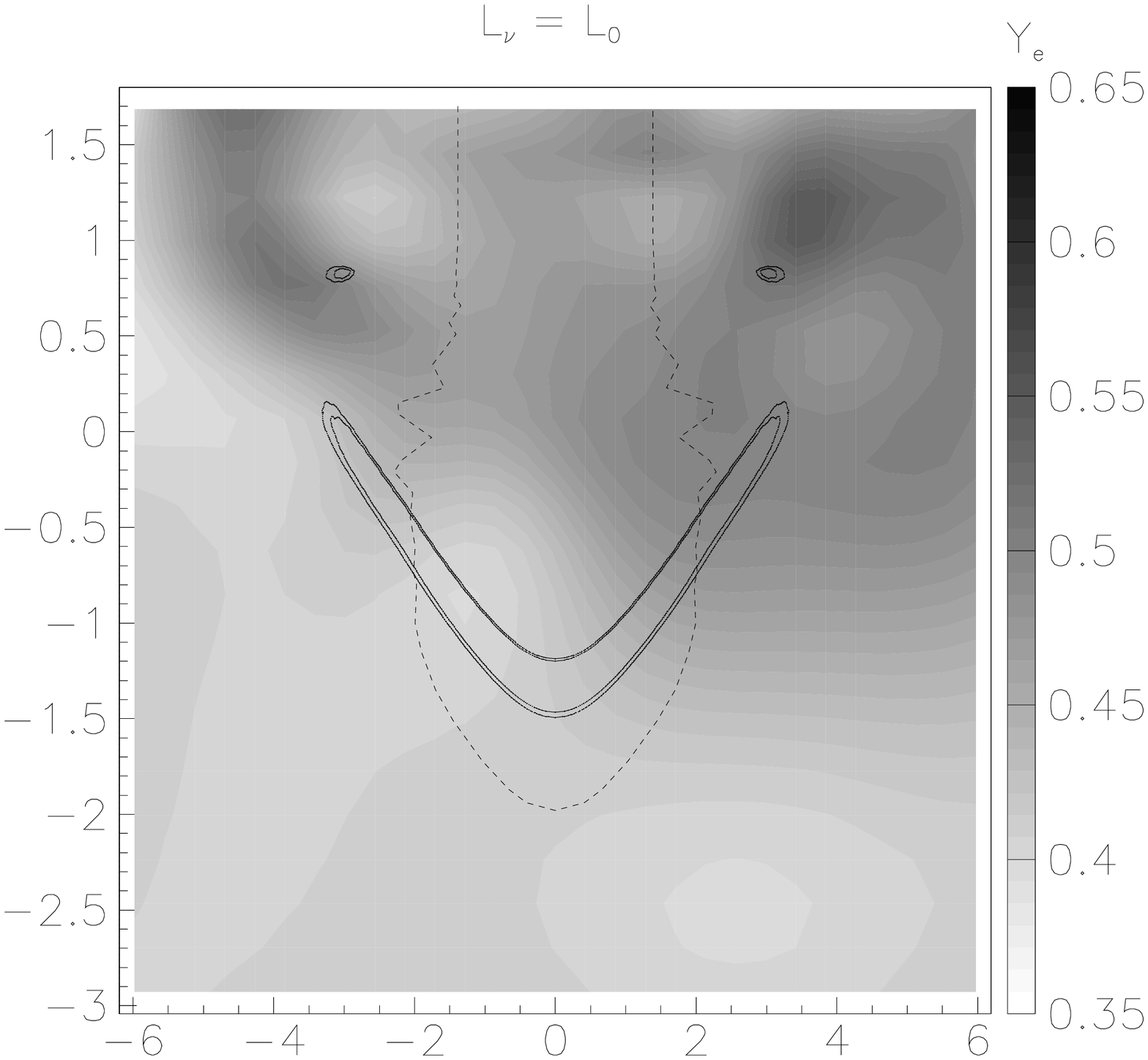}
\includegraphics[width=.328\textwidth]{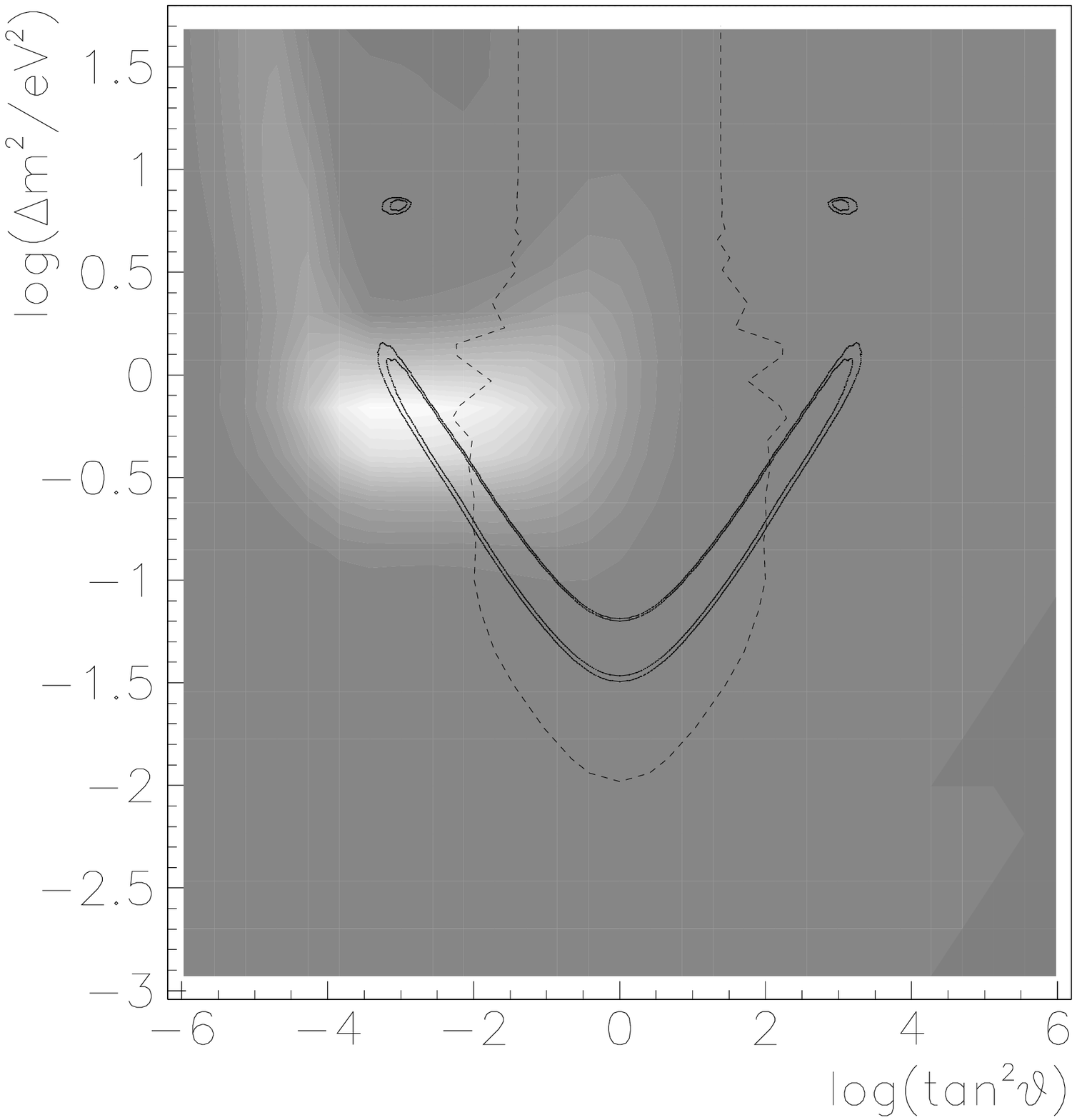}
\includegraphics[width=.328\textwidth]{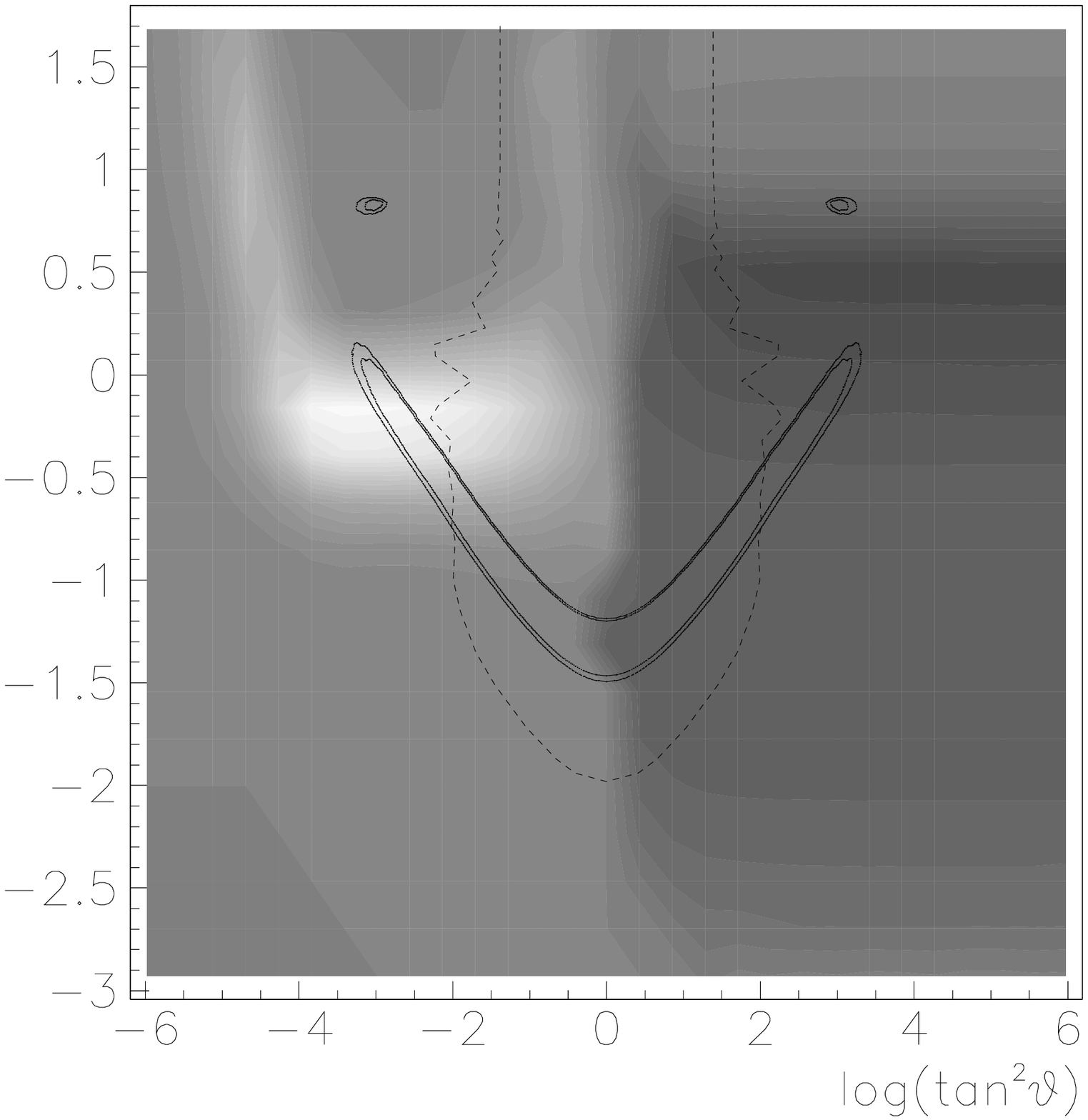}
\includegraphics[width=.328\textwidth]{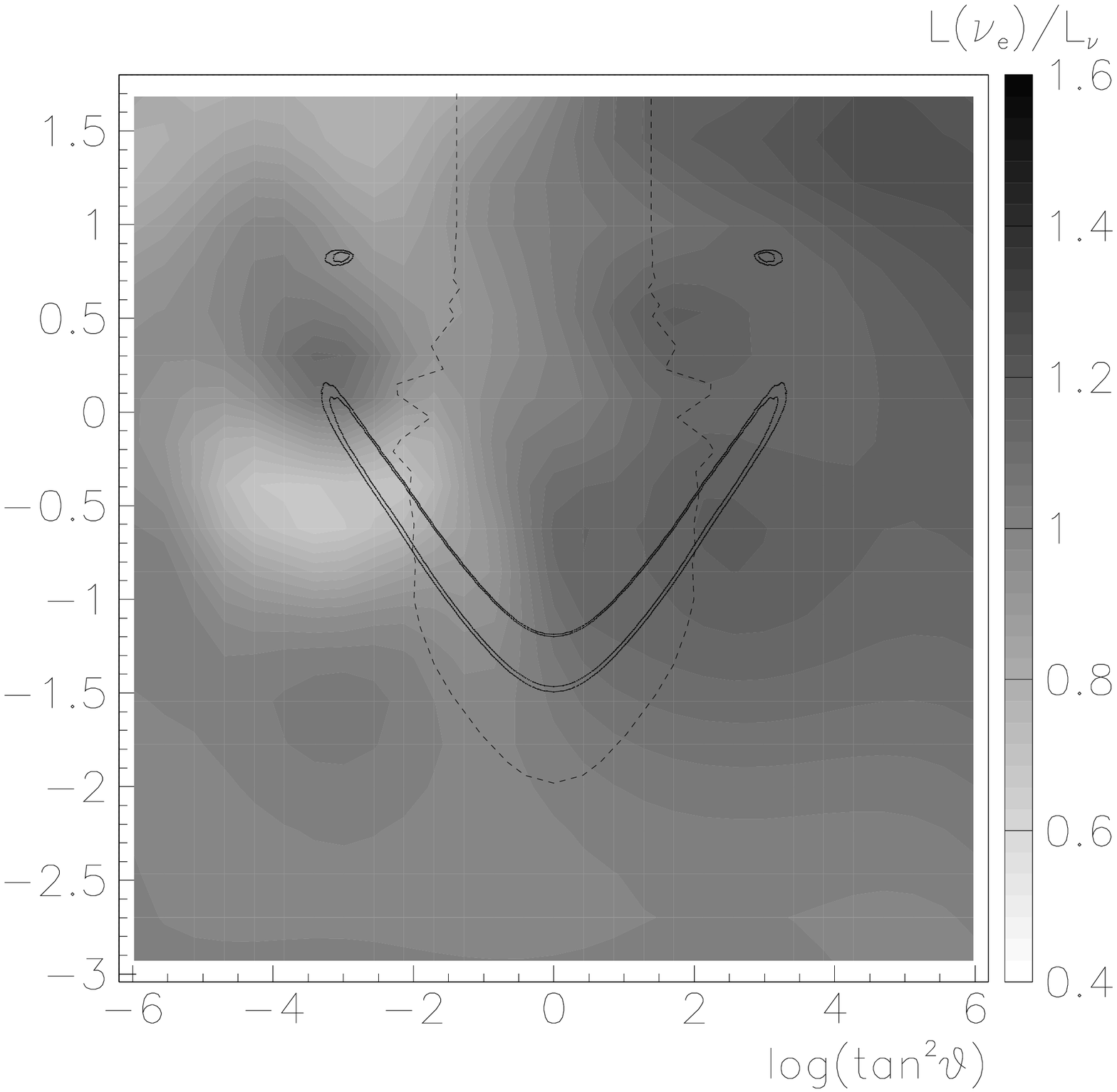}
\end{center}
\caption{Same as Fig.~\ref{contye}, optimized for black-and-white printing.
\label{contyebw}}
\end{figure*}

For our assumed flux spectra we have systematically calculated the
effect of spectral swapping as a function of $\Delta m^2$ and
$\tan^2\theta$. In Fig.~\ref{contye} (for black-and-white printing see
Fig.~\ref{contyebw}) we show our results for the assumed luminosities
$L_\nu=0$ (no neutrino background effects), $L_\nu=0.1$, and
$1\,L_0$. We indicate the region of mixing parameters which is
compatible with the experimental results of LSND and KARMEN2 from a
joint analysis~\cite{Church:2002tc} and the region excluded by
Bugey~\cite{Bugey95}.

In the upper panels we show $Y_e$ at the WFO radius $\simeq30$~km.  In
the absence of neutrino background effects ($L_\nu=0$) our results
agree with those from the previous literature~\cite{Qian:dg}.  For
instance large $\Delta m^2$ and small $\tan^2 \theta$ cause $Y_e >
0.5$, violating the minimal requirement for r-process nucleosynthesis.
However, such regions gradually disappear when the neutrino background
is enhanced, i.e.~neutrino background effects prevent $Y_e$ from
exceeding 0.5, in stark contrast to the previous literature.
Likewise, for an inverted mass situation ($\tan^2\theta>1$) spectral
swapping effects are quite significant even though neutrinos do not
encounter an MSW resonance.

Previous studies of the neutrino background effect used various
approximations~\cite{Pantaleone:1994ns,Qian:wh,Sigl:1994hc}.
Refs.~\cite{Qian:wh,Sigl:1994hc} considered the full set of equations
in an approximate way, but did not include the evolution of
anti-neutrinos. We interpret the fundamental difference between the
previous literature and our results as being caused by the
simultaneous oscillations of neutrinos and anti-neutrinos.

Another intriguing aspect of our results is the behavior of
$L({\nu_e})$ which we show in the lower panels of Fig.~\ref{contye}
for $r=50$~km. It is approximately this region where the protons and
neutrons of the material which would eventually undergo heavy-element
synthesis form alpha particles \cite{alpha1,alpha2,alpha3}.
Therefore, an equal number of protons and neutrons is locked into
alphas, the excess of either of them remaining free.  But an excess of
neutrons can be erased by $\nu_e$ capture if $L({\nu_e})$ is large
enough (``$\alpha$-effect''). One speculative way of reducing
$L({\nu_e})$ invokes oscillations into sterile neutrinos
\cite{McLaughlin:1999pd,Caldwell:1999zk,Fetter:2002xx}.  In our case
of active-active oscillations there is a range of mixing parameters
where $Y_e<0.5$ at the WFO radius, while at larger radii $L({\nu_e})$
is significantly reduced, thus circumventing the $\alpha$-problem.
This happens because the formation of alphas occurs while only
the low-energy $\nu_e$'s are converted, corresponding to the dips in
the $Y_e$ evolution shown in Fig.~\ref{yeevol}. The relevant region
coincides with part of the range allowed by LSND+KARMEN2 and Bugey.
However, this effect persists only for relatively small values of
$L_\nu$.

{\it Summary.}---We find that the impact of neutrino-neutrino
refractive effects in the SN hot bubble region differs markedly from
the established wisdom. Contrary to a naive expectation, the
simultaneous effect of neutrino and anti-neutrino oscillations is
crucial, even if only one of them encounters a resonance. In our
calculation the electron fraction $Y_e$ was never enhanced above $0.5$
when neutrino background effects were included, thus
fulfilling the minimal condition for r-process nucleosynthesis.

If the LSND signature is not due to neutrino conversions and the
active-active oscillation parameters permanently settle in the regions
indicated by solar and atmospheric neutrino conversions, then it is
unlikely that neutrino oscillations influence r-process
nucleosynthesis, always assuming the SN hot bubble region is the
correct site.  There remain interesting oscillation effects at larger
radii where the matter density is smaller~\cite{Schirato:2002tg}, but
we do not expect $\nu$-$\nu$-refraction to play a major role at these
distances. This caveat notwithstanding we find the nonlinear effects
of $\nu$-$\nu$-refraction in the SN hot bubble region a fascinating
topic worth investigating.

{\it Acknowledgments}.---We thank A.~Dighe, G.~Fuller,
M.~Kachelrie\ss, M.~Maltoni, Y.-Z.~Qian, G.~Sigl, and R.~Tom\`as for
helpful comments, and K.~Eitel for sending us the LSND+KARMEN2 data.
This work was partly supported by the Deut\-sche
For\-schungs\-ge\-mein\-schaft under grant No.\ SFB 375 and the ESF
Network Neutrino Astrophysics. SP was supported by a Marie Curie
fellowship of the European Commission under contract
HPMFCT-2000-00445.


\end{document}